\documentclass[aps,showpacs,twocolumn]{revtex4}
\usepackage[dvips]{graphicx}
\usepackage{amssymb}
\usepackage{amsmath}

\newcommand{\btab}{\begin{tabbing}}
\newcommand{\etab}{\end{tabbing}}

\newcommand{\beqn}{\begin{equation}}
\newcommand{\eeqn}{\end{equation}}
\newcommand{\barr}[1]{\begin{array}{#1}}
\newcommand{\earr}{\end{array}}
\newcommand{\beqna}{\begin{eqnarray}}
\newcommand{\eeqna}{\end{eqnarray}}
\newcommand{\btablec}{\begin{table} \begin{center}}
\newcommand{\etablec}{\end{center} \end{table}}

\newcommand{\gapproxeq}{\lower.7ex\hbox{$\;\stackrel{\textstyle>}
{\sim}\;$}}

\begin{document}
\title{On the near-threshold peak structure in the differential cross section of $\phi$-meson photoproduction: indication of a missing resonance with non-negligible strangeness content}
\author{Alvin Kiswandhi}
\author{Shin Nan Yang}
\affiliation{Center for Theoretical Sciences, National Taiwan University,
Taipei 10617, Taiwan \\
Department of Physics, National Taiwan University,
Taipei 10617, Taiwan}
\date{\today}
\begin{abstract}

The details of the analysis of the near-threshold bump structure in
the forward differential cross section of the $\phi$-meson
photoproduction to determine whether it is a signature of a
resonance, together with more extensive results, are presented. The
analysis is carried out in an effective Lagrangian approach which
includes Pomeron and $(\pi, \eta)$ exchanges in the $t$ channel, and
contributions from the $s$- and $u$-channel excitations of a
postulated nucleon resonance. In addition to the differential cross
sections, we use the nine spin-density matrix elements as recently
measured, instead of the $\phi$-meson decay angular distributions
which depend only on six spin-density matrix elements as was done
before, to constrain the resonance parameters. We conclude that
indeed the nonmonotonic behavior, along with the other experimental
data as reported by LEPS, can only be explained with an assumption
of the excitation of a resonance of spin 3/2, as previously
reported. However, both parities of $(\pm)$ can account for the data
equally well with almost identical mass of $2.08\pm 0.04$ GeV and
width of $ 0.501\pm 0.117$ and  $0.570\pm 0.159$ for $3/2^+$ and
$3/2^-$, respectively. The ratio of the helicity amplitudes
$A_{\frac 12}/A_{\frac 32}$ calculated from the resulting coupling
constants differs in sign from that of the known $D_{13}(2080)$.
More experimental data on single and double polarization observables
are needed to resolve the parity. We further find that with an
assumption of large values of the OZI-evading parameters
$x_{\textrm{OZI}} = 12$ for $J^P = 3/2^-$ and $x_{\textrm{OZI}} = 9$
for $J^P = 3/2^+$, the discrepancy between the recent experimental
data on $\omega$-meson photoproduction and theoretical model can be
considerably reduced. We argue that the large value of
$x_{\textrm{OZI}}$ indicates that the postulated resonance contains
non-negligible amount of strangeness content.

\end{abstract}


\maketitle
\section{Introduction}

The $\phi$-meson photoproduction reaction has long been extensively
studied. At high energy, diffractive process dominates and  it can
be well described by $t$-channel Pomeron $(P)$ exchange
\cite{bauer78,donnachie87}. In the low-energy region, the
nondiffractive processes of the pseudoscalar $(\pi, \eta)$-meson
exchanges are also known to contribute \cite{bauer78}. Other
processes, such as nucleon exchange \cite{williams98,oh01}, nucleon
resonances \cite{zhao99,titov03}, second Pomeron exchange,
$t$-channel scalar meson and glueball exchanges
\cite{titov99,titov03}, and $s\bar s$-cluster knockout
\cite{titov97,titov98,oh01} have also been investigated. However,  a
peak in the differential cross sections of $\phi$ photoproduction on
protons at forward angles   around $E_{\gamma}\sim 2.0$ GeV  as
recently observed by the LEPS collaboration \cite{leps05}  cannot be
explained by the processes mentioned above.

Since a bump in the cross sections is often associated with
excitation of resonances, it is then tempting to see if the peak
observed in Ref. \cite{leps05} can be described by a resonance. There
exist previous works studying the effects of resonances in $s$ and
$u$ channels with masses up to 2 GeV \cite{zhao99,titov03}. Ref.
\cite{zhao99} employs $\textrm{SU}(6)\otimes \textrm{O}(3)$ symmetry
within a constituent quark model and included explicitly excited
resonances with quantum numbers $n\leq 2$.  On the other hand, Ref.
\cite{titov03} includes all the known 12 resonances below 2 GeV
listed in Particle Data Group \cite{PDG10}, with coupling constants
determined by available experimental data \cite{besch74,anciant00}
at large momentum transfers. The resonances are found to play
significant roles in the polarization observables. Nevertheless, the
resonances considered, either listed in PDG table or predicted by
some quark model, cannot account for the nonmonotonic behavior as
reported in Ref. \cite{leps05}.

In Ref. \cite{Kiswandhi10}, we have tried to  explore the possibility on
whether such a nonmonotonic behavior   could be explained by a
postulated resonance by fiat  in the neighborhood of observed peak
position. We found that, with an addition of a resonance of spin 3/2
to a background mechanism which consists of Pomeron and $(\pi,
\eta)$-meson exchanges in $t$-channel, not only the  peak in the
forward differential cross section but also the $t$ dependence of
differential cross section (DCS) and $\phi$ meson decay angular
distribution can be well described. Similar attempt was also made in
Ref. \cite{Ozaki09}, where the effect of the $K\Lambda(1520)$ is taken
into account in a coupled-channel analysis. Their results preferred
a resonance of $J^P=1/2^-$.  However, the calculation is marred by a
mistake in the phase of the Pomeron amplitude.

In this paper,  we give the details of our  previous analysis
 \cite{Kiswandhi10} and present  more extensive results of our
calculation. In addition, we employ the new LEPS data \cite{Chang10}
which consist of nine spin-density matrix elements measured at three
different energies to determine the resonance parameters, instead of
the decay angular distributions of the $\phi$ meson, which involve
only six spin-density matrix elements, taken only at two energies
given in Ref. \cite{leps05}, as was done before. The use of a larger data
set with better precision should provide a more stringent constraint
on the model and give rise to more reliable extracted resonance
properties. We also provide an estimation of the strangeness content
of the postulated resonance.

This paper is organized as follows.   The model used in our
analysis, which consists of Pomeron and $(\pi, \eta)$-meson exchanges
in $t$ channel, and a postulated resonance is given in Sec. II. The
extracted resonance parameters, their possible effects in the
polarization observables and $\omega$ photoproduction, as well as an
estimation of the strangeness content of the resonance, are
presented in Sec. III.   The summary is given in Sec. IV.

\section{The Model For $\phi$ Meson Photoproduction}
\label{sec:formalism}
We first introduce the kinematic variables $k$, $p_i$, $q$, and
$p_f$ for the four-momenta of the incoming photon, initial proton,
outgoing $\phi$-meson, and final proton, respectively, with
$s=(k+p_i)^2=(q+p_f)^2$, $t=(q-k)^2 = (p_f-p_i)^2$, and
$u=(p_f-k)^2=(q-p_i)^2$.

\begin{figure}
\includegraphics[width = 1.0\linewidth,angle=0]{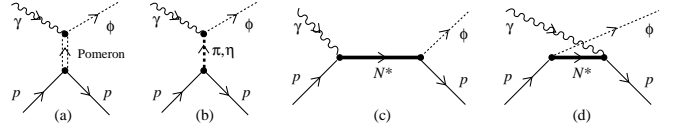}
\caption{Pomeron, $(\pi,\eta)$ exchanges, $s$-, $u$-channel $N^*$ excitation diagrams for $\gamma p \to \phi p$ reaction are labeled (a), (b), (c), and (d), respectively.} 
\label{gammapdiagram}
\end{figure}

We follow the convention of PDG \cite{PDG10} and define the
invariant amplitude $-i{\cal M}$ as related to the $S$-matrix by
\begin{equation}
S_{fi} = \delta_{fi} - i{(2\pi)^4 \delta^{(4)}\left(p_f + q - p_i - k\right)\over (2 E_{{\bf p}_f})^{1/2} (2 E_{{\bf p}_i})^{1/2} (2 E_{\bf q})^{1/2} (2 E_{\bf k})^{1/2}}
{\cal{M}}_{fi},
\end{equation}
with  normalization $\langle p_f | p_i \rangle = (2\pi)^3
\delta^{(3)}({\bf p}_f - {\bf p}_i)$ for free-particle momentum
state and $\bar{u}(p,s)u(p,s) = 2m$ for Dirac spinor with mass $m$.
  In addition to the background mechanism of   Pomeron-exchange, $t$-channel $\pi$-
and $\eta$-exchange, we will postulate the existence of a resonance
by fiat and see whether we could describe the data of LEPS
\cite{leps05,Chang10}. We can then write the full amplitude ${\cal
M}$ as
\begin{eqnarray}
{\cal M}_{fi} ={\cal M}_P + {\cal M}_{\pi + \eta} + {\cal M}_{N^*},
\label{amp}
\end{eqnarray}
as shown in Fig.~\ref{gammapdiagram}, where ${\cal M}_{N^*}$
contains both $s$- and $u$-channel contributions. The unpolarized
differential cross section is related to the invariant amplitude by
\begin{equation}
\frac{d\sigma}{dt}=\frac{1}{64\pi
s|{\bf{k}}_{cm}|^2}\frac{1}{4}\sum_{\lambda_N,\lambda_{\gamma}}\sum_{\lambda_{N'},\lambda_{\phi}}|{\cal
M}_{fi}|^2,
\end{equation}
with ${\bf{k}}_{cm}$ is the photon three momentum in the
center-of-mass (CM) frame and $\lambda_N$, $\lambda_{N'}$,
$\lambda_{\gamma}$ and $\lambda_{\phi}$  denote the helicities of
the initial proton, final proton, incoming photon, and outgoing
$\phi$-meson, respectively.

\subsection{Pomeron exchange}
\label{sec:pomeron}
Following Refs.~\cite{titov03,titov07}, we can easily write down the
Pomeron-exchange amplitude of Fig.~\ref{gammapdiagram}(d) ,
\begin{eqnarray}
{\cal M}_P &=& -\bar{u}(p_f,\lambda_{N'})M(s,t)\Gamma^{\mu \nu} u(p_i,\lambda_N) \nonumber \\
&\times&\varepsilon^*_{\mu}(q,\lambda_{\phi})\varepsilon_{\nu}(k,\lambda_{\gamma}),
\label{ampa}
\end{eqnarray}
where $\varepsilon_{\mu}(q,\lambda_{\phi})$ and
$\varepsilon_{\nu}(k,\lambda_{\gamma})$ are the polarization vectors
of the $\phi$-meson and photon with  $\lambda_{\phi}$ and
$\lambda_{\gamma}$, respectively, and
$u(p_i,\lambda_N)$[$u(p_f,\lambda_{N'})$] is the Dirac spinor of the
nucleon with momentum $p_i$($p_f$) and helicity
$\lambda_N$($\lambda_{N'}$). The transition operator $\Gamma^{\mu
\nu}$ in Eq. (\ref{ampa}) is
\begin{eqnarray}
\Gamma^{\mu \nu} &=& \left(g^{\mu \nu}-\frac{q^{\mu} q^{\nu}}{q^2}\right)
\not\!k - \left(k^{\mu} - \frac{k \cdot q q^{\mu}}{q^2}\right)
\gamma^{\nu} \nonumber \\
&-& \left(\gamma^{\mu}-\frac{\not\!q
q^{\mu}}{q^2}\right)\left[q^{\nu} - \frac{k \cdot
q(p^{\nu}_i+p^{\nu}_f)}{k \cdot (p_i + p_f)}\right].
\end{eqnarray}
The scalar function $M(s,t)$ is described by the Reggeon
parametrization,
\begin{eqnarray}
M(s,t) &=& C_P
F_1(t)F_2(t)\frac{1}{s}\left(\frac{s-s_{th}}{s_0}\right)^{\alpha_P(t)} \nonumber \\
&\times&\text{exp}\left[-\frac{i\pi}{2}\alpha_P(t)\right],
\end{eqnarray}
where the Pomeron trajectory is taken to be $\alpha_P(t) = 1.08 +
0.25 t$ and $s_0=(m_N+m_\phi)^2$.   $F_1(t)$, the isoscalar form
factor of the nucleon and $F_2(t)$, the form factor of the
$\phi$-photon-Pomeron coupling are given as
\cite{titov03,donnachie87},
\begin{eqnarray}
F_1(t)&=& \frac{4m^2_N-a_N^2t}{(4m^2_N-t)(1-t/t_0)^2},\\
F_2(t)&=& \frac{2\mu^2_0}{(1-t/m^2_{\phi})(2\mu^2_0+m^2_{\phi}-t)},
\end{eqnarray}
with $\mu^2_0 = 1.1$ $\textrm{GeV}^2$, $a_N^2 = 2.8$, and $t_0 =
0.7$ $\textrm{GeV}^2$.

\begin{figure}[htbp]
\vspace{-0.5cm}

\includegraphics[width=0.95\linewidth,angle=0]{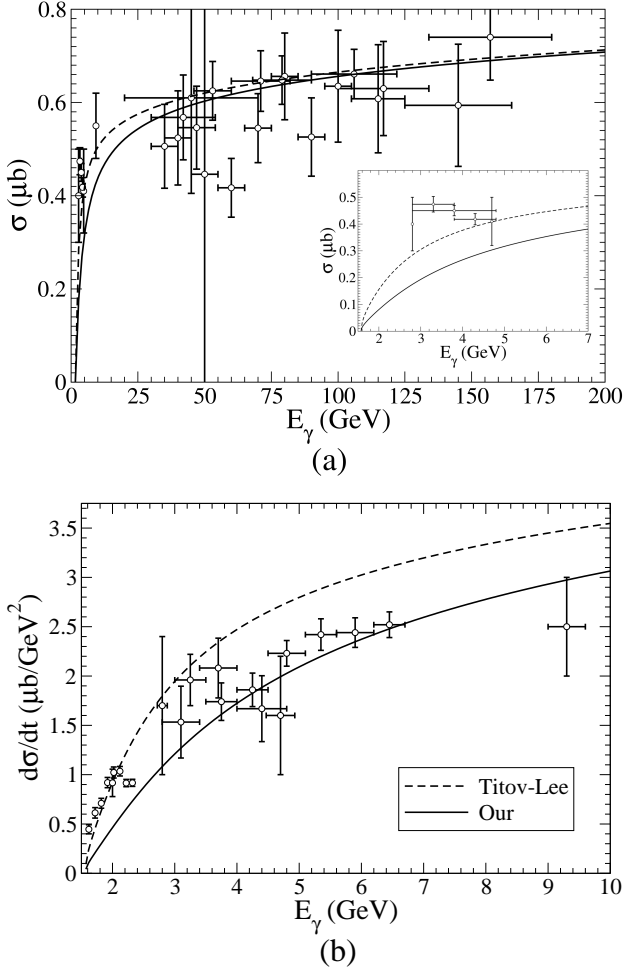}\\
\vspace{0.3cm}

\includegraphics[width=0.95\linewidth,angle=0]{pomeron_compare_low.eps}
\caption{(a) Total cross sections of $\phi$ photoproduction as a
function of photon lab energy $E_\gamma$. The inset gives an
enlarged view for the region with $E_\gamma \leq$ 7 GeV.
(b) Differential cross sections of $\phi$ photoproduction at forward
angle as a function of photon lab energy $E_\gamma$. Results of this
work (Our) and Ref.~\cite{titov03} (Titov-Lee) are drawn as full and
dashed lines, respectively. Data is taken from Ref.~\cite{durham}.}
\label{pomeron_comparison}
\end{figure}

In this study, we follow Ref.~\cite{titov03} by choosing the
strength factor $C_P = 3.65$, which is obtained by fitting to the
total cross sections data at high energy, as shown in the upper
figure in Fig.~\ref{pomeron_comparison}, where the inset show the
enlarged view of the region for $E_\gamma \leq$ 7 GeV.  We include a
threshold factor $s_{th}$ as was done in
Refs.~\cite{williams98,titov03} in order to get a better agreement
with  experimental data near the threshold region. If  $s_{th}=0$ is
chosen as done in Ref.~\cite{titov03}, a problem arises. Namely, the
results for forward differential cross sections would overestimate
the experimental data \cite{durham} by about $20 \%$ as seen in the
lower figure of Fig.~\ref{pomeron_comparison} around $E_\gamma = 6$
GeV. Since pomeron properties and behaviors at lower energies are
not well-established,  we adjust this parameter to fit the
experimental data on the differential cross sections around
$E_\gamma = 6$ GeV. $E_\gamma = 6$ GeV is chosen because at this
energy, one can reasonably expect that all other contributions from
hadronic intermediate states would become negligible and only
pomeron contributes. Furthermore, around this energy,  experimental
data
  are quite reliable in that they have relatively small error
bars and   rise steadily without much fluctuation. These give us
confidence to match the pomeron contribution to the experimental
data at this energy by fixing $s_{th} = 1.3$ ${\textrm{GeV}^2}$.

\subsection{$\pi$ and $\eta$-meson exchanges}

The amplitudes for the $\pi$ and $\eta$ exchanges in $t$ channel,
Fig.~\ref{gammapdiagram}(c) in a straightforward manner
\cite{titov97,titov98} and are given by
\begin{eqnarray}
{\cal M}_{\pi + \eta} &=& \frac{-eg_{\gamma\phi \pi}g_{\pi
NN}F^2_{\pi}(t)}{m_{\phi}}\bar{u}(p_f,\lambda_{N'})\gamma_5\frac{\varepsilon^{\mu\nu\rho\sigma}q_{\mu}k_{\rho}}{t-m^2_{\pi}} \nonumber \\
&\times& u(p_i,\lambda_N)\varepsilon^*_{\nu}(q,\lambda_{\phi})\varepsilon_{\sigma}(k,\lambda_{\gamma})+
\nonumber \\
&-& \frac{eg_{\gamma\phi \eta}g_{\eta
NN}F^2_{\eta}(t)}{m_{\phi}}\bar{u}(p_f,\lambda_{N'})\gamma_5\frac{\varepsilon^{\mu\nu\rho\sigma}q_{\mu}k_{\rho}}{t-m^2_{\eta}} \nonumber \\
&\times& u(p_i,\lambda_N)\varepsilon^*_{\nu}(q,\lambda_{\phi})\varepsilon_{\sigma}(k,\lambda_{\gamma}),
\label{ampb}
\end{eqnarray}
with the coupling constants $g_{\pi NN}$, $g_{\gamma \phi \pi}$, and
$g_{\gamma \phi \eta}$, as well as the form factors $F_{\pi}(t)$ and
$F_{\eta}(t)$ for the virtually exchanged mesons at the $MNN$ and
$\gamma \phi M$ ($M=\pi,\eta$) vertices, respectively, are taken to
be the same as in Ref.~\cite{titov07}. We choose $g_{\eta NN} =
1.12$~\cite{wentai} and $\Lambda_{\pi} = \Lambda_{\eta} = 1.2$ GeV
which are slightly different with the values given in
Ref.~\cite{titov07}. The choice of the cut-off parameter
$\Lambda_{\pi}=1.2$ GeV would lead to a $F_{\pi}(t)$ which agrees
well with the $\pi NN$ form factors as obtained in the
meson-exchange $\pi N$ model Ref. \cite{piN} in  the region of 
$-0.5 < t  < 0 \textrm{ GeV}^2$ where most of the data considered  in this present
work lie.

\subsection{Excitation of a baryon resonance}

The  Feynman diagrams with an $N^*$ in the intermediate state in $s$-
and $u$-channel   are shown in Fig.~\ref{gammapdiagram}(a) and (b).
To evaluate the invariant amplitudes involving $N^*$, we use the
following interaction Lagrangians. For the coupling of spin-$1/2$
and $3/2$ resonances to $\gamma N$, we choose the commonly used
interaction Lagrangians~\cite{wentai,feu,pascalutsa}
\begin{eqnarray}
{\cal L}_{\gamma N N^*}^{1/2^\pm} &=& e g^{(2)}_{\gamma N N^*}
\bar{\psi}_N \Gamma^\pm \sigma_{\mu \nu} F^{\mu \nu} \psi_{N^*} +
\textrm{h.c.}, \\ \label{photononehalf}
{\cal L}_{\gamma N N^*}^{3/2^\pm} &=& i e g^{(1)}_{\gamma N N^*} \bar{\psi}_{N} \Gamma^\pm \left(\partial^\mu\psi_{N^*}^\nu\right) \tilde{F}_{\mu \nu} \nonumber \\
&+& eg^{(2)}_{\gamma N N^*} \bar{\psi}_{N} \Gamma^\pm \gamma^5 \left(\partial^{\mu} \psi_{N^*}^\nu \right)F_{\mu \nu} + \textrm{h.c.}, \label{photonthreehalf}
\end{eqnarray}
where $F_{\mu \nu}= \partial_{\mu}A_{\nu} - \partial_{\nu}A_{\mu}$
is the electromagnetic field tensor, and $\sigma_{\mu \nu} =
\frac{i}{2}(\gamma_{\mu}\gamma_{\nu}-\gamma_{\nu} \gamma_{\mu})$.
Also, $\tilde{F}_{\mu \nu} = {1 \over 2}
\epsilon_{\mu\nu\alpha\beta} F^{\alpha\beta}$ denotes the dual
electromagnetic field tensor with $\epsilon^{0123} = +1$. The
operator $\Gamma^\pm$ are given by $\Gamma^+=1$ and
$\Gamma^-=\gamma_5$. For the $\phi N N^*$ interaction Lagrangians,
we have
\begin{eqnarray}
{\cal L}_{\phi N N^*}^{1/2^\pm} &=& g_{\phi N N^*}^{(1)} \bar{\psi}_{N} \Gamma^\pm \gamma^\mu \psi_{N^*} \phi_{\mu} \nonumber \\
&+&g_{\phi N N^*}^{(2)}\bar{\psi}_{N} \Gamma^\pm
\sigma_{\mu \nu} G^{\mu \nu} \psi_{N^*} + \textrm{h.c.}, \\ \label{phiNNstaronehalf}
{\cal L}_{\phi N N^*}^{3/2^\pm} &=& ig^{(1)}_{\phi N N^*}
\bar{\psi}_{N} \Gamma^\pm \left(\partial^\mu\psi_{N^*}^\nu\right) \tilde{G}_{\mu \nu} \nonumber \\
&+& g^{(2)}_{\phi N N^*} \bar{\psi}_{N} \Gamma^\pm \gamma^5
\left(\partial^{\mu} \psi_{N^*}^\nu \right)G_{\mu \nu} \nonumber \\
&+& ig^{(3)}_{\phi N N^*} \bar{\psi}_{N} \Gamma^\pm \gamma^5  \gamma_\alpha
\left(\partial^\alpha \psi^\nu_{N^*} - \partial^\nu \psi^\alpha_{N^*}\right)
\left(\partial^\mu G_{\mu\nu}\right) \nonumber \\
&+& \textrm{h.c.}, \label{phiNNstar}
\end{eqnarray}
where   $G^{\mu \nu}$ is defined as $G^{\mu \nu} =
\partial^{\mu} \phi^{\nu}-\partial^{\nu} \phi^{\mu}$ with $\phi^{\mu}$ the field of $\phi$-meson. The dual field tensor $\tilde{G}_{\mu \nu}$ is again
defined in the same way as its electromagnetic counterpart with
$F^{\alpha\beta} \rightarrow G^{\alpha\beta}$. Notice that we could
have chosen to describe the $\gamma N N^*$ in the same way as we
describe the $\phi N N^*$ interactions. However,   current
conservation consideration fixes $g^{(1)}_{\gamma N N^*}$ for $J^P =
1/2^\pm$ resonances to be zero.  In addition, the term proportional
to $g^{(3)}_{\gamma N N^*}$ in  the Lagrangian densities of Eq.
(\ref{phiNNstar}) vanishes in the case of real photon. With the
Lagrangians given in Eqs. (\ref{photononehalf}-\ref{phiNNstar}), the
full invariant amplitude of $s$ and $u$ channels can readily be
written down straightforwardly by following the Feynman rules.

The form factor for the vertices used in the $s$- and $u$-channel
diagrams, $F_{N^*}(p^2)$, is taken to be similar as in
Ref.~\cite{piN}
\begin{equation}
F_{N^*}(p^2)=\frac{\Lambda^{4}}{\Lambda^{4} + (p^2-M^2_{N^*})^2},
\end{equation}
with $\Lambda$ is the cut-off parameter for the virtual $N^*$. In this work, we choose $\Lambda = 1.2$ GeV for all resonances.
The spin-$1/2$ $N^*$ propagator can be written in a Breit-Wigner form as
\begin{equation}
G^{(1/2)}(p)=\frac{i(\not\! p
+M_{N^*})}{p^2-M^2_{N^*}+iM_{N^*}\Gamma_{N^*}},
\end{equation}
with $\Gamma_{N^*}$ the total decay width of $N^*$. The
Rarita-Schwinger propagator is used for the spin-$3/2$ $N^*$
\begin{eqnarray}
G_{\mu\nu}^{(3/2)}(p)&=&\frac{i(\not\! p+M_{N^*})}{p^2-M^2_{N^*}+iM_{N^*}\Gamma_{N^*}}\left[-g_{\mu\nu} + {1 \over 3} \gamma_\mu \gamma_\nu \right.\nonumber \\
&-& \left. {1 \over 3 M_{N^*}} \left(p_\mu \gamma_\nu - p_\nu \gamma_\mu\right) + {2 \over 3 M_{N^*}^2} p_\mu p_\nu\right].\nonumber \\
\end{eqnarray}
Because $u<0$, we take $\Gamma_{N^*} = 0 $ MeV for the propagator in
the $u$ channel.

It should, however, be stressed that, we do not know the value of
the coupling constants $g_{\phi N N^*}$ and $g_{\gamma N N^*}$, as
our calculations are done in the tree level. Therefore, in present
calculation, we show the values of $g_{\gamma N N^*}g_{\phi N N^*}$
obtained by fitting the experimental data.

\begin{table*}
\caption{\label{table} The $N^*$ parameters for $J^P = 3/2^\pm$
resonances together with their errors obtained by HESSE method of MINUIT package.}

\begin{center}
\begin{tabular}{|c||c|c|}
\hline
                                                                & $J^P = 3/2^+$                        & $J^P = 3/2^-$                         \\
\hline\hline
$M_{N^*}$(GeV)                                                  & 2.08 $\pm$  0.04                     & 2.08 $\pm$ 0.04                       \\
$\Gamma_{N^*}$(GeV)                                             &  0.501 $\pm$ 0.117                   & 0.570 $\pm$ 0.159                     \\
\hline
$e{g}_{\gamma N N^*}^{(1)}{g}_{\phi N N^*}^{(1)}$               & $\quad$ 0.003 $\pm$ 0.009 $\quad$    &$\quad$ $-$0.205 $\pm$ 0.083 $\quad$   \\

$\quad$$e{g}_{\gamma N N^*}^{(1)}{g}_{\phi N N^*}^{(2)}$$\quad$ & $-$0.084  $\pm$ 0.057                & $-$0.025 $\pm$ 0.017                  \\

$e{g}_{\gamma N N^*}^{(1)}{g}_{\phi N N^*}^{(3)}$               & 0.025  $\pm$ 0.076                   & $-$0.033 $\pm$ 0.017                  \\

$e{g}_{\gamma N N^*}^{(2)}{g}_{\phi N N^*}^{(1)}$               & 0.002 $\pm$ 0.006                    & $-$0.266 $\pm$ 0.127                  \\

$e{g}_{\gamma N N^*}^{(2)}{g}_{\phi N N^*}^{(2)}$               & $-$0.048 $\pm$ 0.047                 & $-$0.033 $\pm$ 0.032                  \\

$e{g}_{\gamma N N^*}^{(2)}{g}_{\phi N N^*}^{(3)}$               & 0.014 $\pm$ 0.040                    & $-$0.043 $\pm$ 0.032                  \\
\hline
$\chi^2/N$                                                      &0.891                                 & 0.821                                 \\

\hline
\end{tabular}
\end{center} \label{tab:nstar}
\end{table*}

\section{Results and Discussions}
 With the model presented
in Sec. \ref{sec:formalism}, one can easily obtain the full
amplitude of $\gamma p \to \phi p$ reaction and calculate   the
scattering observables straightforwardly with a specific assignment
of spin-parity of the resonance. Since the peak appears to lie close
to the $\phi N$ threshold, only the lower partial waves are
important and we shall consider only $J^P = 1/2^\pm, 3/2^\pm$ as the
possible candidates for the spin-parity assignment of the resonance.
\begin{figure}[h]
\vspace{0.5cm}

\includegraphics[width=0.90\linewidth,angle=0]{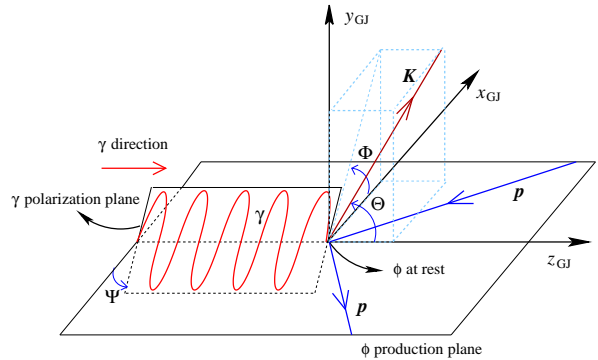}

\caption{(Color online) The $\phi$ photoproduction in Gottfried-Jackson system.}
\label{GJ_frame}
\end{figure}
 In this work, we fit our model simultaneously to differential cross section
at forward angle as a function of photon energy and differential
cross sections dependence on $t$ at eight photon energies reported
in Ref. \cite{leps05}, as well as to nine spin-density matrix elements
$\rho^\alpha_{ij}$ \cite{schillingnpb15397} as a a function of $t$
at three photon energies \cite{Chang10}. We use the
Gottfried-Jackson system, in which $\phi$-meson is at rest, as
depicted in Fig. \ref{GJ_frame}, to analyze the spin-density matrix
elements. The $z_{GJ}$ axis is taken to be along the incoming photon
momentum while $y_{GJ}$ axis is taken to be along $\bf{p}_f \times
\bf{p}_i$ direction, with $\bf{p}_f $ and $  \bf{p}_i$ the
three-momentum of final and initial proton, respectively. The
$x_{GJ}$ axis is chosen to form a right-handed coordinate system.

Notice that in our previous work \cite{Kiswandhi10},    decay
angular distributions $W$, instead of the spin-density matrix
elements $\rho^\alpha_{ij}$ were used in the data set to which we
fit our model parameters. Even though the decay angular
distributions are also functions of spin-density matrix elements
(SDME), they depend only on only six out of a total of nine SDMEs.
Furthermore, the decay angular distributions used in
Ref. \cite{Kiswandhi10}, as presented in Ref. \cite{leps05},   are taken only
at two photon energies and averaged over $t$, while the nine SDMEs
used in this work as presented in Ref. \cite{Chang10} are taken at three
photon energies and are functions of $t$.    We expect that the
larger set of data considered in this work would  provide a more
stringent constraint on our model and results.

In  tree-level approximation, only products like $g_{\gamma N
N^*}g_{\phi N N^*}$ enter. The other parameters in our model are the
resonance mass and width. They are determined with the use of MINUIT
by fitting to the data measured at SPring8 \cite{leps05,Chang10} as
described in the previous paragraphs.

We find that, with assignments of spin-parity $J^P = 1/2^\pm$ for the resonance,
the nonmonotonic behavior in the forward differential cross-section near threshold
can be explained only with considerably stronger resonance contributions. As a result,
the differential cross section as a function of $t$, as well as spin-density matrix elements, would be in disagreement with the experimental data \cite{leps05,Chang10}.
The resulting $\chi^2/N$ from such fit will be around $5 \sim 9$, which is definitely far above those obtained by fitting using $J^P = 3/2^\pm$ resonances, 
as seen in Table \ref{tab:nstar}.
Therefore, we conclude that spin-1/2 resonance cannot fit the experimental data. It is worthwhile to note that in the
constituent quark model of Refs.~\cite{capstick1,capstick2},
spin-1/2 resonances are also not predicted to be of significant
contribution at around $E_\gamma = 2$ GeV. Our results seem to be in
line with their prediction.

On the other hand, we find that the experimental data can be well
described with a spin-parity assignment of either $J^P = 3/2^-$ or
$J^P = 3/2^+$ for the postulated resonance.
In the following, we first present our model predictions for the
differential cross sections, spin-density matrix elements, and decay
angular distributions and compare them with the data for both the
cases with $3/2^-$ and $3/2^+$ resonances. We then present an
analysis on the composition of the bump structure. After that, we
proceed by predicting the effect on $\omega N$ channel and
estimating the strangeness content of the resonance. Lastly, we
present also some predictions on the polarization observables.

\subsection{Differential cross sections, spin-density matrix
elements, and decay angular distributions}

 The quality of the agreement between data and model predictions for
 both spin-parity assignments, i.e., $3/2^\pm$, is similar even
though the resulting $\chi^2$ value is slightly smaller for the case
$J^P = 3/2^-$ as seen in Table~\ref{tab:nstar} where the values of
the products of $g_{\gamma N N^*}g_{\phi N N^*}$ are also presented.
The obtained values for the mass and width for   $3/2^- $ and
$3/2^+$ resonances are very close, i.e., (mass, width) of (2.08,
0.570) and (2.08, 0.501) GeV, respectively, however,  the products
of the coupling constants are quite different.

\begin{figure}[b]

\includegraphics[width=0.95\linewidth,angle=0]{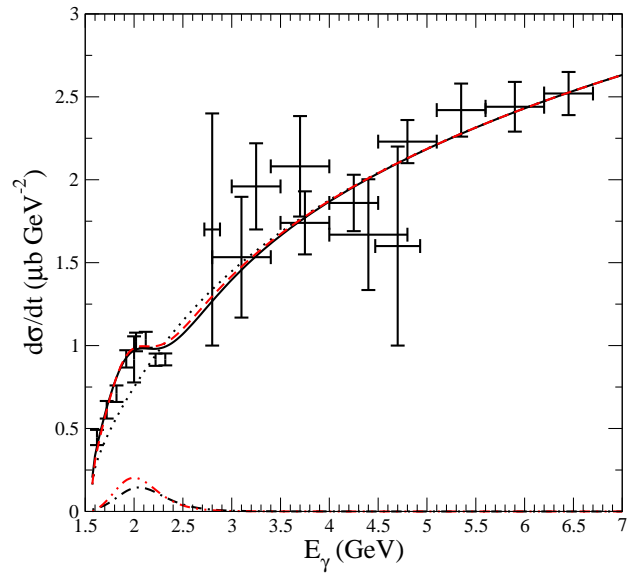}

\caption{(Color online) The results obtained by employing a $J^P = 3/2^\pm$
resonances in our model for differential cross section of $\gamma p
\to \phi p$ at forward direction as a function of the incoming
photon energy $E_\gamma$. Data are taken from Ref.~\cite{durham}.
The dotted lines represent the background which includes pomeron-
and meson-exchange contributions only. The full and dashed lines are
the total contributions including $J^P = 3/2^-$ and $J^P = 3/2^+$
resonances, respectively. The dash-dotted and dash-dot-dotted lines
denote the resonant $s$- and $u$-channel contributions of $J^P =
3/2^-$ and $J^P = 3/2^+$ resonances, respectively.} \label{DCS_s}
\end{figure}

\begin{figure*}[htbp]

\includegraphics[width=0.95\linewidth,angle=0]{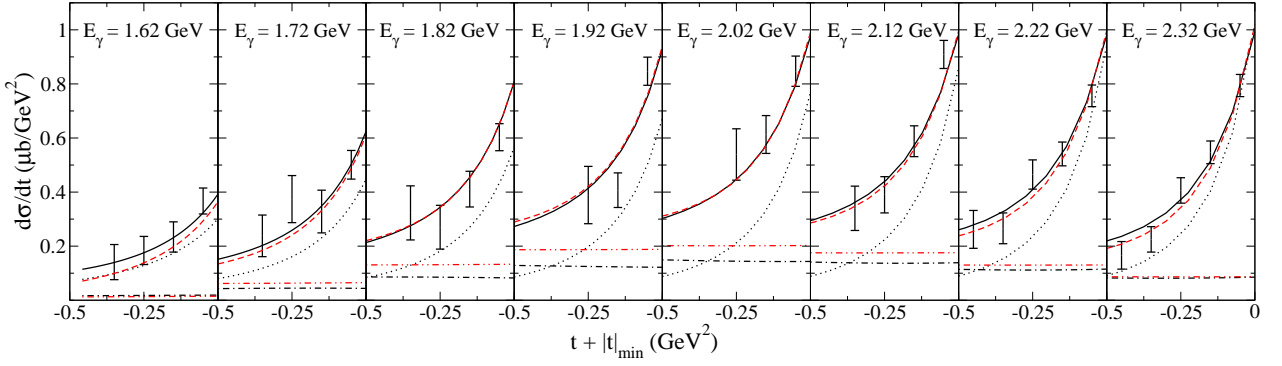}
\vspace{0.cm}

\caption{(Color online) The results obtained for differential cross section of
$\gamma p \to \phi p$ as a function of $t + |t|_{\textrm{min}}$ at
eight different photon energies $E_\gamma$, as given inside each plot. Data is taken from Ref.~\cite{durham}.
Notation is as in Fig. \ref{DCS_s}.} \label{DCS_t}
\end{figure*}

\begin{figure}[b]

\includegraphics[width=1.0\linewidth,angle=0]{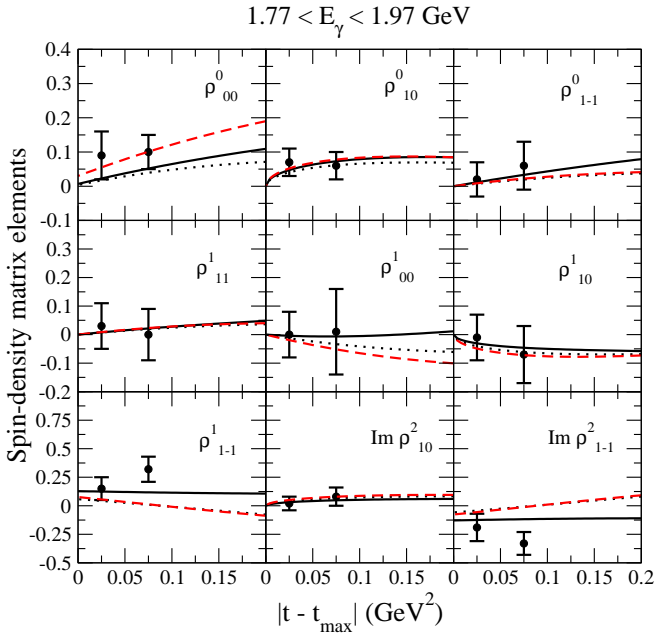}

\caption{(Color online) Spin-density matrix elements $\rho^\alpha_{ij}$ in Gottfried-Jackson system as
a function of $t$ at $1.77<E_\gamma<1.97$ GeV. Notation is as in
Fig. \ref{DCS_s}.} \label{GJ_187}
\end{figure}

\begin{figure}[b]

\includegraphics[width=1.0\linewidth,angle=0]{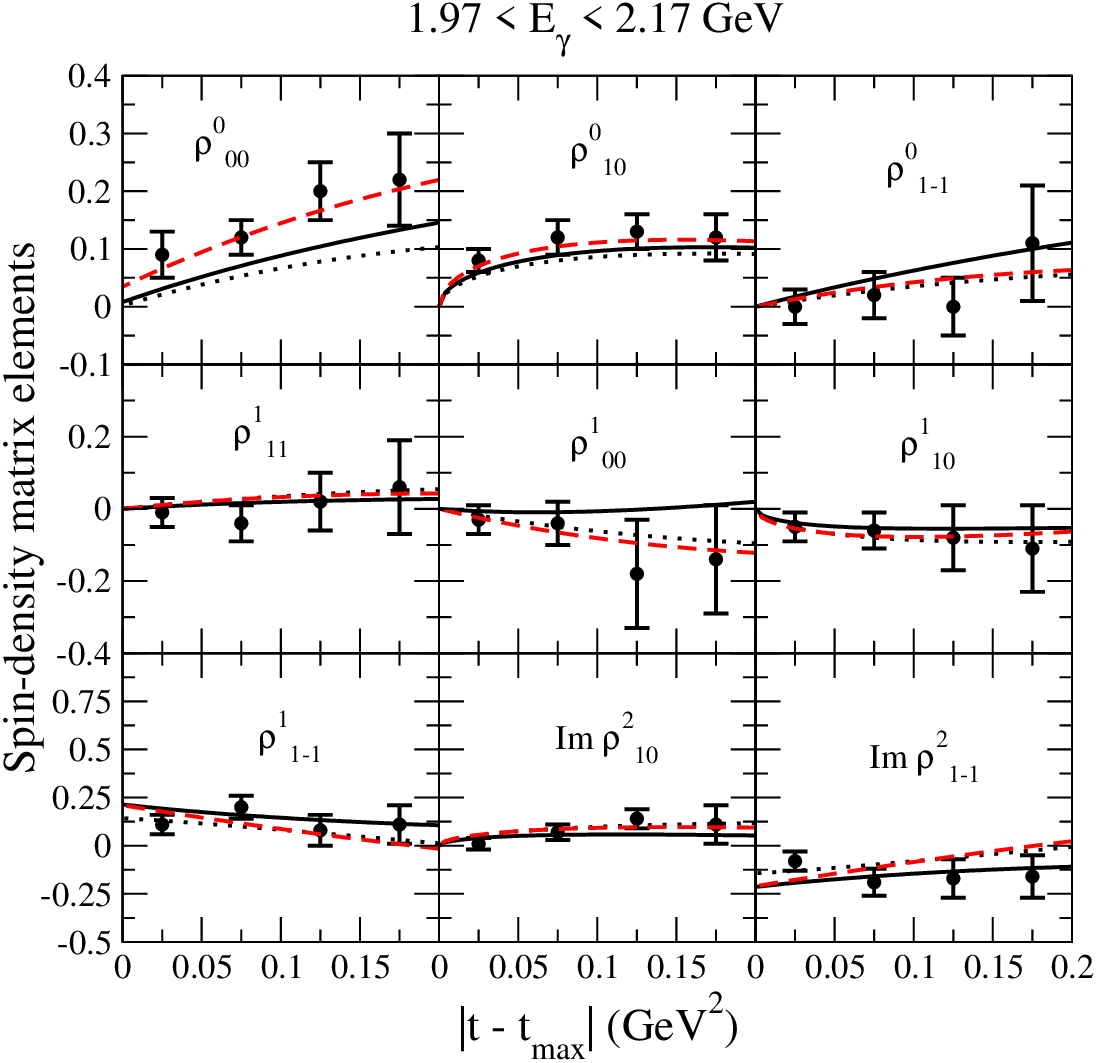}

\caption{(Color online) Spin-density matrix elements $\rho^\alpha_{ij}$ in Gottfried-Jackson system as
a function of $t$ at $1.97<E_\gamma<2.17$ GeV. Notation is as in
Fig. \ref{DCS_s}.} \label{GJ_207}
\end{figure}

\begin{figure}[htbp]

\includegraphics[width=1.0\linewidth,angle=0]{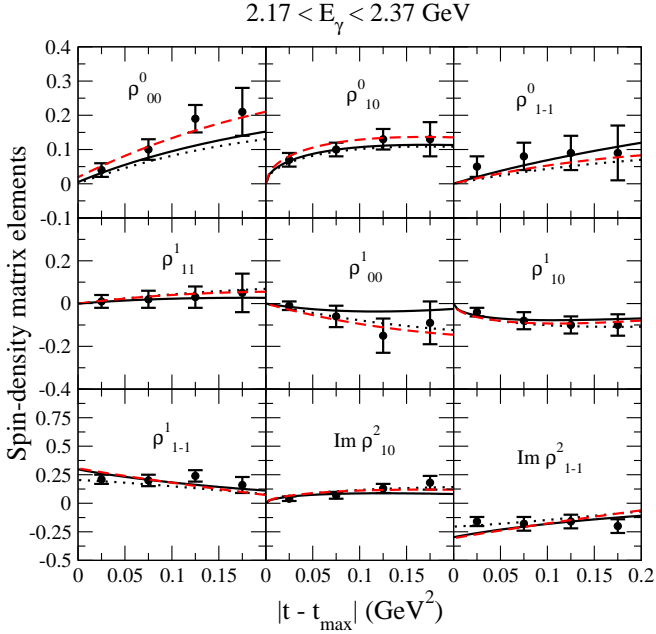}

\caption{(Color online) Spin-density matrix elements $\rho^\alpha_{ij}$ in Gottfried-Jackson system as
a function of $t$ at $2.17<E_\gamma<2.37$ GeV. Notation is as in
Fig. \ref{DCS_s}.} \label{GJ_227}
\end{figure}

The results of our best fit, as compared to the data of
Refs. \cite{leps05,Chang10}, are shown in Fig. \ref{DCS_s}$-$\ref{GJ_227}.
The dotted lines represent the contributions of the background of
pomeron plus $(\pi, \eta)$ exchanges and the solid and dashed curves
correspond to the full model predictions including a resonance of
$3/2^-$ and $3/2^+$, respectively. In Figs. \ref{DCS_s} and
\ref{DCS_t}, the forward differential cross section as function of
energy, where a bump is observed, and the differential cross section
as function of $t$ are shown, respectively. The contribution of the
resonance alone is also shown therein with dash-dotted and
dash-dot-dotted lines corresponding to $3/2^-$ and $3/2^+$,
respectively. We see that besides producing a bump in the forward
differential cross section, the resonance reduces the discrepancy
between predictions of the background mechanisms and the data
substantially in the $t$ dependence of the differential cross
sections.

\begin{figure}[b]

\includegraphics[width=1.0\linewidth,angle=0]{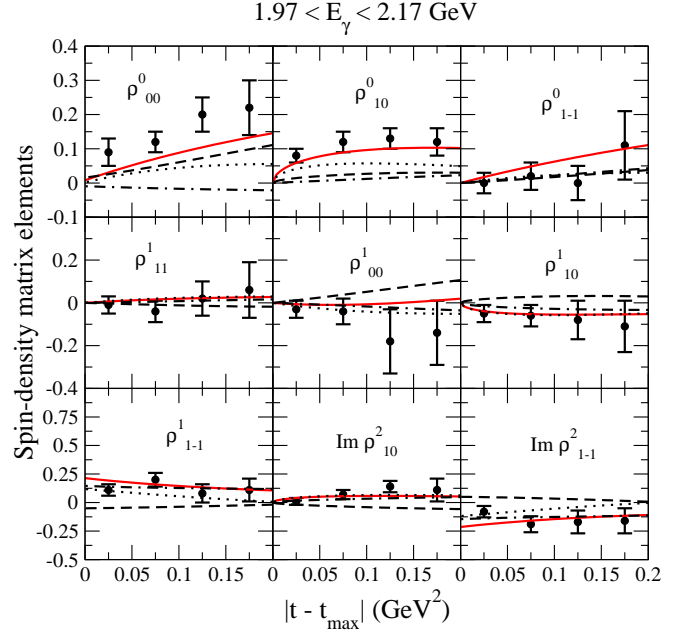}

\caption{(Color online) Detail of the composition of spin-density matrix elements $\rho^\alpha_{ij}$ in Gottfried-Jackson system as a function of $t$ at $1.97<E_\gamma<2.17$ GeV for a $J^P = 3/2^-$ resonance. Note that we have different notation here.
The full, dotted, dashed, and dash-dotted lines are the total, nonresonant, resonant, and interference between nonresonant and resonant contributions, respectively.} \label{GJ_207_min}
\end{figure}

\begin{figure}[htbp]
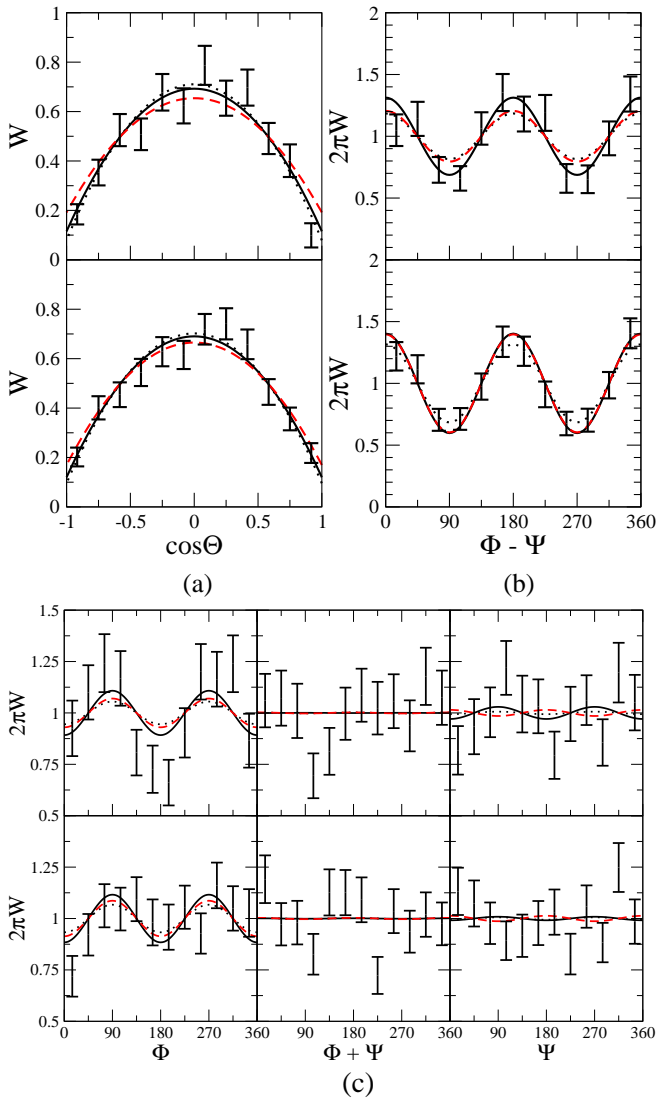


\includegraphics[width=1.0\linewidth,angle=0]{W_theta.eps}\\

\includegraphics[width=1.0\linewidth,angle=0]{W_phi.eps}\\

\caption{(Color online) Decay angular distributions (a) $W(\cos{\Theta})$, (b) $W(\Phi-\Psi)$, (c) $W(\Phi)$, $W(\Phi+\Psi)$, and $W(\Psi)$, at photon LAB energies $1.97-2.17$ GeV
(upper panels) and $2.17-2.37$ GeV (lower panels) within the range of $|t-t_\textrm{max}| \le 0.2 \textrm{ GeV}^2$. Notation is as in Fig. \ref{DCS_s}.} \label{W}
\end{figure}

In Figs. \ref{GJ_187}-\ref{GJ_227}, our model predictions for the
SDME's in the three energy regions of $1.77 < E_\gamma < 1.97$ GeV,
$1.97 < E_\gamma < 2.17$ GeV, and $2.17 < E_\gamma < 2.37$ GeV are shown
together with the data of Ref. \cite{Chang10}. It is seen that in some
cases, e.g.,  $\rho^0_{10}$,  $\rho^1_{10}$, $\rho^2_{10}$ in $1.97 <
E_\gamma < 2.17$ GeV, the nonresonant contribution alone already
describes well the data and the resonance contributions are small.
However, there are several cases that a $3/2^+$ resonance  is indeed
quite helpful to bridge the difference between background
contribution and the data, especially for $\rho^0_{00}$ in all
energy regions, though its corrections are in the wrong direction
for $\rho^1_{1,-1}$ and $\rho^2_{1,-1}$ in the region of $1.77 <
E_\gamma < 1.97$ GeV. The effect of a $3/2^-$ resonance is in general
less conspicuous than that of a $3/2^+$ resonance.

The decay angular distributions $W(\cos{\Theta})$,
$W(\Phi-\Psi)$, $W(\Phi)$, $W(\Phi+\Psi)$, and $W(\Psi)$ in
Gottfried-Jackson frame depend on six SDMEs via the following relations,
\begin{eqnarray}
W(\cos \Theta) &=& {3 \over 2}\left[{1 \over 2}(1 - \rho^0_{00})\sin^2 \Theta + \rho^0_{00} \cos^2 \Theta \right], \nonumber \\
W(\Phi) &=& {1 \over 2\pi}(1 - 2 \textrm{Re} \rho^0_{1-1} \cos 2\Phi), \nonumber \\
W(\Phi - \Psi) &=& {1 \over 2\pi} \{1 + 2 P_\gamma (\rho^1_{1-1} - \textrm{Im}\rho^2_{1-1}) \nonumber \\
&\times& \cos \left[2 (\Phi - \Psi)\right]\},\nonumber
\end{eqnarray}
\begin{eqnarray}
W(\Phi + \Psi) &=& {1 \over 2\pi} \{1 + 2 P_\gamma (\rho^1_{1-1} + \textrm{Im}\rho^2_{1-1}) \nonumber \\
&\times& \cos \left[2 (\Phi + \Psi)\right]\}, \nonumber \\
W(\Psi) &=& {1 \over 2\pi} \left[1 - P_\gamma (2\rho^1_{11} +
\rho^1_{00})\cos 2\Psi\right], \label{W_and_rho}
\end{eqnarray}
where the angles $\Theta$, $\Phi$, and $\Psi$ are illustrated in
Fig. \ref{GJ_frame}. Here, they are measured at two different energy 
bins $1.97 < E_\gamma < 2.17$ GeV and $2.17 < E_\gamma < 2.37$ GeV within the range of
$|t-t_\textrm{max}| \le 0.2 \textrm{GeV}^2$
($t_\textrm{max}=-|t|_\textrm{min}$). In our work, they are calculated at
the midpoint of each energy bin $E_\gamma$ by weighing them with the
differential cross section as a function of $t$ \beqna
&&W(E_\gamma, \Theta, \Phi, \Psi) = \nonumber \\
&&\frac{\int_{t_\textrm{max}-0.2}^{t_\textrm{max}} dt
\left[d\sigma(E_\gamma, t)/dt\right] W(E_\gamma, t, \Theta, \Phi,
\Psi)} {\int_{t_\textrm{max}-0.2}^{t_\textrm{max}} dt
\left[d\sigma(E_\gamma, t)/dt\right]}. \eeqna It is important to
note that it is misleading to conclude that the effect of the
resonance, be it $3/2^+$ or $3/2^-$, is insignificant in most cases.
Fig. \ref{GJ_207_min} shows the detail of the composition of the
spin-density matrix elements as a function of $t$ at
$1.97<E_\gamma<2.17$ GeV for a $J^P = 3/2^-$ where the full, dotted,
dashed, and dash-dotted lines are the total, nonresonant, resonant,
and interference between nonresonant and resonant contributions,
respectively. It is obvious that the resonant contributions are not
negligible compared to the nonresonant ones. However, the
interference contributions are also roughly of the same strength as
those of the resonance, and in many cases, of the opposite signs.
This would cause the total contributions to come mainly from the
nonresonant contributions only. However, it should be emphasized
again, that the resonant contributions are not negligible.

Based on the similarities in their masses and spin-parities, one
might wonder whether the $3/2^-$ resonance found here can be
identified as the $D_{13}(2080)$ as listed in PDG \cite{PDG10}. The
coupling constants given Table \ref{tab:nstar} can be used to
calculate the ratio of the helicity amplitudes $A_{1/2}$ and
$A_{3/2}$. However, we cannot determine their magnitudes since we
have only the products of the coupling constants  $\gamma NN^*$ and
$\phi NN^*$. We obtain a value of {$A_{1/2}/A_{3/2}=1.05$}, while
it is $-1.18$ for $D_{13}(2080)$. Although their magnitudes are
quite similar, they differ by a sign and we conclude that the
resonance postulated here, if exists, cannot be identified with
$D_{13}(2080)$.

\begin{figure}[htbp]
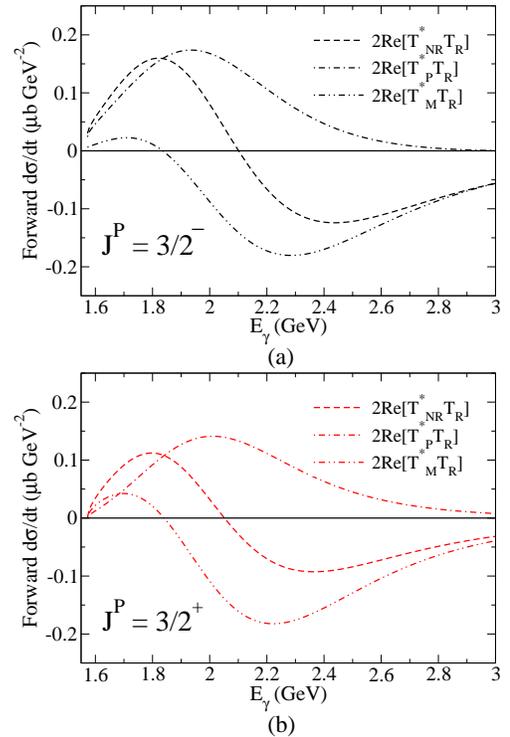


\includegraphics[width=0.75\linewidth,angle=0]{interference_2_minus.eps}\\

\includegraphics[width=0.75\linewidth,angle=0]{interference_2_plus.eps}

\caption{(Color online) Decomposition of the contribution of the
interference between the resonant and nonresonant amplitudes, $
2\textrm{Re}(T_\textit{NR} T_R^*)$, to the the differential cross
section at forward angle as a function of incident photon energy
$E_\gamma$ for a resonance with (a) $J^P = 3/2^-$ and (b) $J^P =
3/2^+$. The  dot-dashed, and dot-dot-dashed lines denote
the interferences of the resonant amplitude with the pomeron and
meson-exchange amplitudes, $ 2\textrm{Re}(T_\textit{P} T_R^*)$ and
  and $ 2\textrm{Re}(T_\textit{M}
T_R^*)$, respectively, while the dashed lines represent the  sum $
2\textrm{Re}(T_\textit{NR} T_R^*)$.} \label{check}
\end{figure}


\subsection{Analysis on the composition of the bump structure}

Our results for the forward differential cross sections of the $J^P
= 3/2^\pm$  in Fig. \ref{DCS_s} indicate constructive and
destructive interferences of nonresonant and resonant amplitudes
below and above the peak, respectively. Since the nonresonant
amplitude is dominated overwhelmingly by Pomeron amplitude, which is
almost completely imaginary, it seems to imply that the
sign-changing component of the resonant part must then be also
imaginary. However, it is well-known that the imaginary part of the
resonant amplitude is sign definite while the sign-changing
component of a resonant amplitude is real .

In order to understand this, let us decompose the forward
differential cross section ($\propto |T|^2$) into its nonresonant,
resonant, and interference terms
\begin{equation}
|T|^2 = |T_\textit{NR}|^2 + |T_R|^2 + 2 \textrm{Re}(T_\textit{NR}
T_R^*),
\end{equation}
in which $T$, $T_\textit{NR}$, and $T_R$ are the total, nonresonant,
and resonant amplitudes, respectively. We also have
\begin{equation}
T_\textit{NR} = T_P + T_M,
\end{equation}
where $T_P$ and $T_M$ are the Pomeron and meson-exchange amplitude,
respectively.
 The interference term between the resonant and
nonresonant amplitude $ \textrm{Re}(T_\textit{NR} T_R^*)$ can be further
decomposed into a sum of interference terms between the resonant and
pomeron amplitudes $ \textrm{Re}(T_\textit{P} T_R^*)$, and the
resonant and meson-exchange amplitudes $ \textrm{Re}(T_\textit{M}
T_R^*)$, respectively. This decomposition is shown in the Fig.
\ref{check}, where the dot-dashed and dot-dot-dashed lines
represent $ \textrm{Re}(T_\textit{P} T_R^*)$ and $
\textrm{Re}(T_\textit{M} T_R^*)$, respectively, with dashed lines
denoting their sum $ \textrm{Re}(T_\textit{NR} T_R^*)$. It is seen
that for both $J^P = 3/2^\pm$ resonances,   $
\textrm{Re}(T_\textit{P} T_R^*)$ is always positive, while $2
\textrm{Re}(T_\textit{M} T_R^*)$ turns negative at around $E_\gamma
= 1.82$ GeV  and becomes comparable in size to $
\textrm{Re}(T_\textit{P} T_R^*)$ such that the sum  $
\textrm{Re}(T_\textit{NR} T_R^*)$ eventually changes sign. We
conclude that the meson-exchange mechanisms are indeed crucial in
producing the peaking behavior observed in $\phi$-photoproduction
reaction.

\subsection{Effects of the postulated resonance in the $\omega N$ channel}

\begin{figure*}[htbp]
\vspace{0.cm}

\includegraphics[width=0.75\linewidth,angle=0]{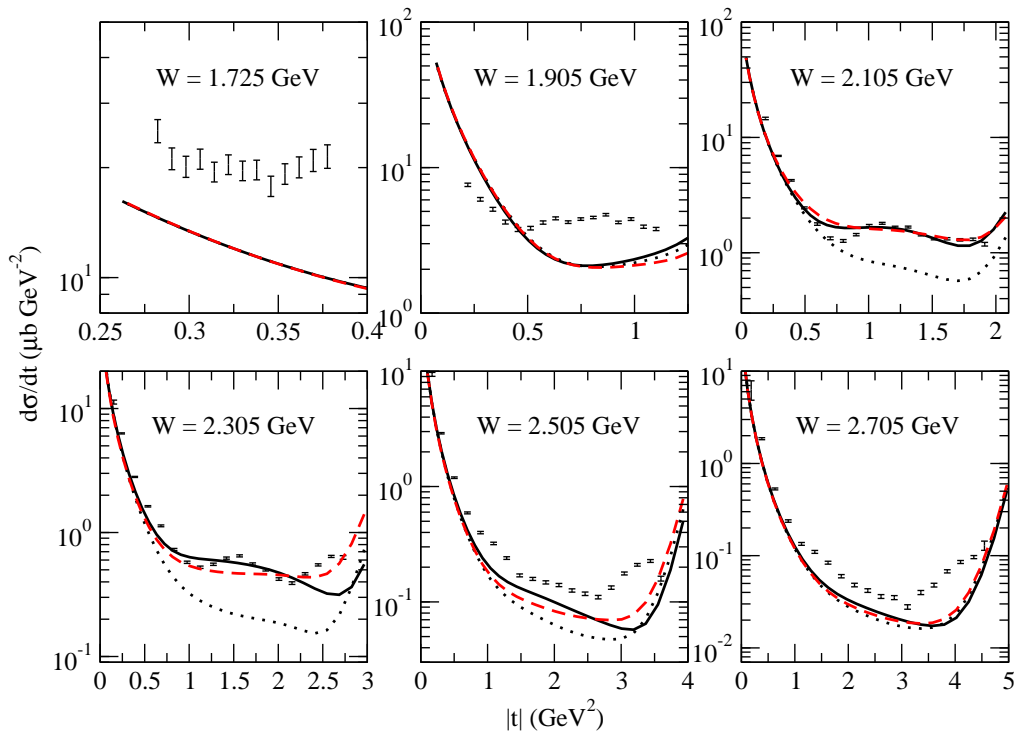}\\
\vspace{0.cm}

\caption{(Color online) Predictions of the effects of the postulated resonances to the differential cross section of $\omega$ photoproduction at CM energies $W$, as given 
inside each plot. Data is from Ref. \cite{M_Williams}. Notation is as in Fig. \ref{DCS_s}.} \label{omega}
\end{figure*}

We  further study the possible effect of this postulated resonance
in the $\omega N$ channel. The conventional "minimal"
parametrization relating $\phi NN^*$ and $\omega NN^*$ is
\begin{eqnarray}
g_{\phi N N^*} = -\tan \Delta \theta_V x_{\text{OZI}} g_{\omega N N^*}, \label{coupligrelation}
\end{eqnarray}
with $\Delta \theta_V \simeq 3.7^\circ$ corresponding to the
deviation from the ideal $\phi-\omega$ mixing angle. Here,
$x_{\text{OZI}}$ is called the OZI-evading parameter and the larger
value of $x_{\text{OZI}}$ would indicate larger strangeness content
of the resonance.

For the present purpose, we choose the $\omega$ photoproduction
model of Ref.~\cite{oh02} which includes the nucleon resonances
predicted by Refs.~\cite{capstick1, capstick2}.  In
Fig.~\ref{omega}, one sees that the prediction of this model for the
$t$-dependence of differential cross section at $W = 2.105$ GeV,
given in dotted line,  exhibits substantial discrepancy with the
most recent experimental data \cite{M_Williams} for $|t| > 0.75$
GeV$^2$. With the addition of resonance postulated here  with
$x_{\text{OZI}}=12 (9)$ for $J^P = 3/2^- (J^P = 3/2^+)$, we see that
the differential cross section at $W = 2.105$ and $2.305$ GeV,  as
denoted by  the solid black (dashed red) line in Fig.~\ref{omega},
can be reproduced with roughly the correct strength.  At the  other
energies, the improvement is much less noticeable because they are
far from the energy of the resonance. The large values of
$x_{\text{OZI}}$ would imply that the resonances we propose here
might contain  considerable amount of strangeness contents, an issue
we now turn to in the next subsection.

\subsection{Strangeness content of the postulated resonance}
\begin{figure}[b]
\includegraphics[width = 1.0\linewidth,angle=0]{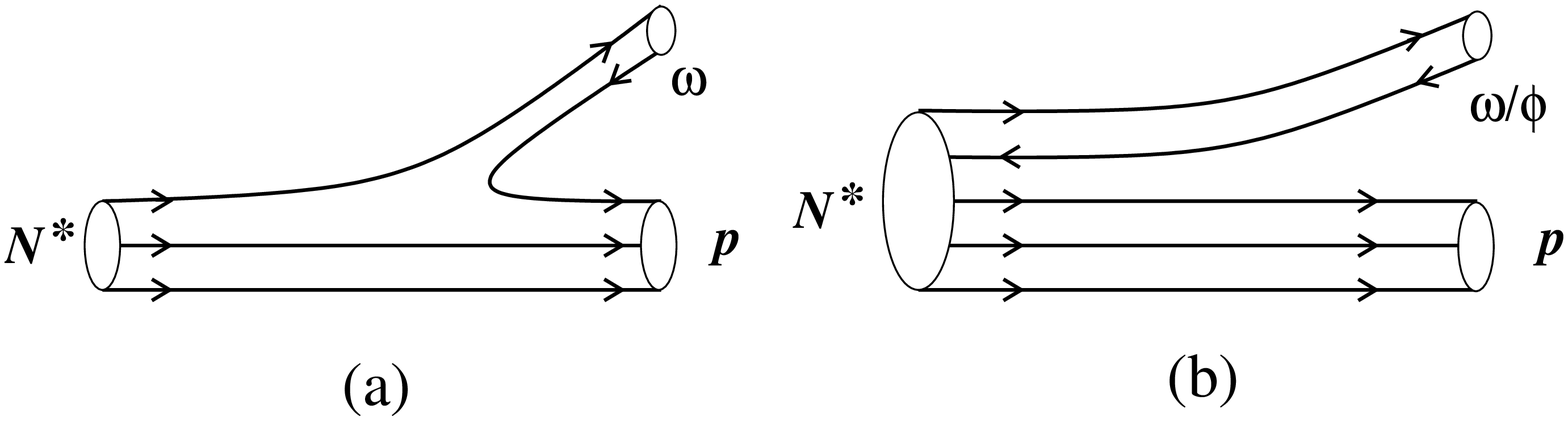}
\caption{Quark-flow diagrams of the processes that correspond to the
amplitudes (a) $\mathcal{M}_3$ and (b) $\mathcal{M}_5$.}
\label{M3M5}
\end{figure}
The resonance proposed here appears to have a large OZI evasion
parameter $x_{\text{OZI}}$ which would lead one to ask whether this
is reasonable. In this section, we will estimate the strangeness
content of the resonance. We can write, for the wavefunction of the
resonance \cite{Ellis88}, \beqn |N^*\rangle = x |uud\rangle + z_u
|uudu\bar u\rangle + z_d |uudd\bar d\rangle + z_s |uuds\bar
s\rangle, \eeqn where $x$ is real but $z_u$, $z_d$, and $z_s$ are
all complex and $|x|^2 + |z_u|^2 + |z_d|^2 + |z_s|^2 = 1$. Let us
define \beqn Z_{id} \equiv \frac{\mathcal{M}(N^* \to
\phi_{id}N)}{\mathcal{M}(N^* \to \omega_{id}N)}, \eeqn where
$\mathcal{M}(N^* \to V_{id}N)$ is the amplitude of the decay of
$N^*$ to $V_{id}N$ where the subscript {\it "id"} denotes that the
vector meson $V$ is in its ideal state. For example, $\phi_{id}$
consists of pure $s \bar s$ with no $u \bar u$ of $d \bar d$
mixture.

We can obtain $Z_{id}$ experimentally from \beqn Z_{id} =
\frac{Z_{\textrm{phys}} + \tan \Delta \theta_V}{1 -
Z_{\textrm{phys}} \tan \Delta \theta_V}, \label{Z_relation} \eeqn
where \beqn Z_{\textrm{phys}} \equiv \frac{\mathcal{M}(N^* \to \phi
N)}{\mathcal{M}(N^* \to \omega N)}, \eeqn is defined for the
physical particles $\phi$ and $\omega$. Here, $Z_{\textrm{phys}}$
can be estimated from $g_{\phi N N^*}/g_{\omega N N^*} = -
x_{\text{OZI}} \tan \Delta \theta_V$. Notice that $Z_{id} = 0$ when
$x_{\text{OZI}} = 1$ which corresponds to the case of ordinary OZI
violation arising from an $\omega\phi$ mixing without the presence
of strangeness content in the resonance $N^*$. By using the values
of $x_{\text{OZI}}$ for the resonance  found in this work, the
values for their $Z_{\textrm{phys}}$ can also be calculated from Eq.
(\ref{coupligrelation}). Therefore, employing Eq. (\ref{Z_relation})
above, we can obtain the values of $Z_{id}$, which are found to be
$-0.68$ and $-0.50$ for $J^P = 3/2^-$ and $J^P = 3/2^+$ resonances,
respectively.

\begin{figure}[htbp]
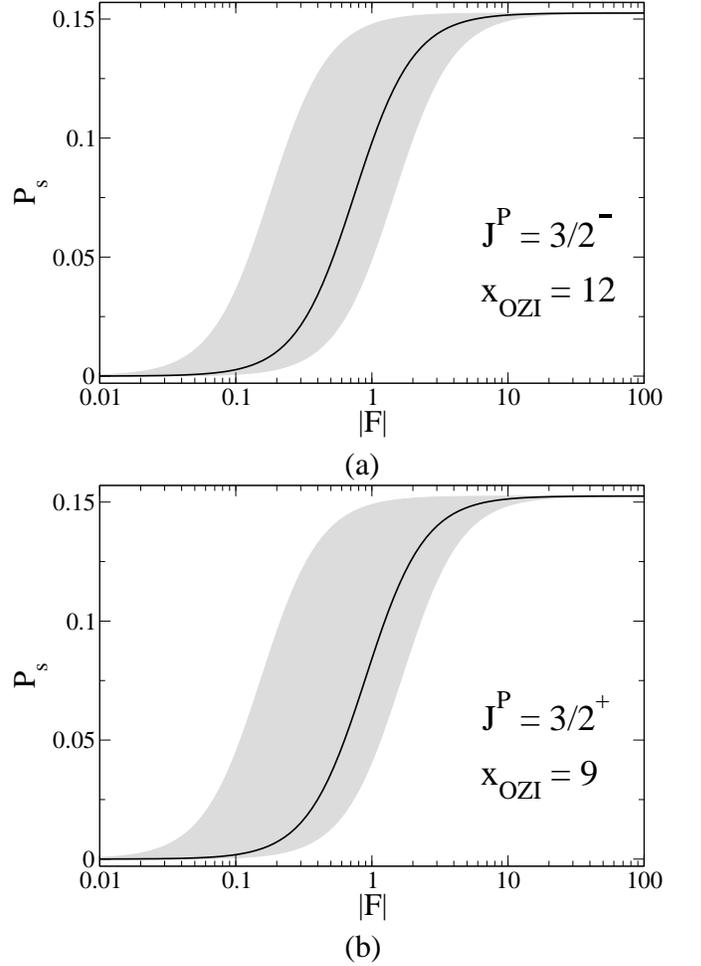

\includegraphics[width = 1.0\linewidth,angle=0]{strange_bar_minus.eps}
\includegraphics[width = 1.0\linewidth,angle=0]{strange_bar_plus.eps}
\caption{Strangeness content of the resonances $P_s$ for $x_{\text{OZI}} = 12$ and $x_{\text{OZI}} = 9$ corresponding to (a) $J^P = 3/2^-$ and (b) $J^P = 3/2^+$ resonances, respectively, as functions of $|F|$. The shaded areas show the $95\%$ probability range after the phases are randomly varied. The solid lines are the median.}
\label{strange}
\end{figure}

Within the constituent quark model,  $Z_{id}$ is related to the
amplitudes $\mathcal{M}_3$ and $\mathcal{M}_5$ corresponding to the
processes depicted in Fig. \ref{M3M5}(a) and (b), respectively, as
followings   \beqna
\mathcal{M}(N^* \to \phi_{id}N) &=& z_s \mathcal{M}_5, \nonumber \\
\mathcal{M}(N^* \to \omega_{id}N) &=& x \mathcal{M}_3 +
\frac{1}{\sqrt{2}}\left(z_u + z_d\right)\mathcal{M}_5. \eeqna Let us
now write $z_q \equiv \delta_q a_q$, where $q = u, d, s$, which
separates the phase factor $\delta_q=e^{i\theta_q}$ of phase
$\theta_q$ and the magnitude $a_q=|z_q|$ of the amplitude. We
further introduce $c_u \equiv a_u/a_s$ and $c_d \equiv a_d/a_s$.
After substituting, we have \beqna Z_{id} &=& \frac{z_s
\mathcal{M}_5}{x \mathcal{M}_3 + \frac{1}{\sqrt{2}}\left(\delta_s^*
\delta_u c_u + \delta_s^* \delta_d c_d \right)z_s\mathcal{M}_5},
\eeqna which leads to   the probability of the strangeness content
\beqn P_s \equiv |z_s|^2 = \frac{|Z_{id}|^2|F|^2}{(1 + c_u^2 +
c_d^2)|Z_{id}|^2|F|^2 +  {|N|^2} }, \eeqn where \beqna N &\equiv&
1 - \frac{1}{\sqrt{2}}Z_{id}\left(\delta_s^*\delta_u c_u + \delta_s^*\delta_d c_d\right), \label{N} \\
F&\equiv&\frac{M_3}{M_5}. \eeqna

It is seen that the strangeness probability $P_s$ depends on
$\mathcal{M}_{3,5}$ and $z_q$'s in a complicated way. The problem
here is to make an educated estimate of it.  Here, we first follow
Ref. \cite{brodsky} to assume that $P_{u,d}/P_s = (m_s/m_{u,d})^2$.
It leads to $c_u = c_d = m_s/m_{u,d}$ where $m_s$ and   $m_{u,d}$
will be taken as 0.5 and 0.3  GeV, respectively  To proceed, we
further assume the ratio   $|F|$ between the reaction amplitudes
$\mathcal{M}_3$ and $\mathcal{M}_5$ to lie within the range of 0.01
and 100, i.e.,
  $|\mathcal{M}_3| = (0.01 \sim 100 )|\mathcal{M}_5|$ and find the possible
range  of $P_s$. For a fixed value of $|F|$, we randomly vary the
phase factors $\delta_q$'s to give $P_s$. The results, within $95\%$
probability, are given by the shaded area in Fig. \ref{strange}
while the median values are denoted by the solid lines.

  Notice that, for a fixed value of  $|F|$, the lower bound of $P_s$ is given by
$|N|_{\textrm{max}} = 1 + \frac{1}{\sqrt{2}}|Z_{id}|\left(c_u +
c_d\right)$, that is, when all the phases are such that the second
term in $N$ of Eq. (\ref{N}), interferes constructively with the
first. The lower bound of  $P_s$ goes to zero like
$|Z_{id}/N_\textrm{max}|^2|F|^2$ as $|F|$ approaches zero and approaches
$1/(1 + c_u^2 + c_d^2)=0.153$ when $|F|$ grows larger. When
$|N|_{\textrm{min}} = 0$, the upper bound of $P_s=1/(1 + c_u^2 +
c_d^2)=0.153$  is reached. Note that with the values of $Z_{id}$ of
the resonance given above, the condition $|N|_{\textrm{min}} = 0$
can indeed be met with some combinations of the phase factors
$\delta_q$'s. We point out that   $P_s=1/(1 + c_u^2 + c_d^2)=0.153$,
would correspond to a resonance with $100\%$ five-quark content,
namely, a pentaquark state.

A rather broad range of $P_s$ also reflects the situation faced in
the efforts to determine the strangeness content in the proton,
which is stable and can be more directly studied. Recent studies
give estimates ranging from $0.025 - 0.058\%$ \cite{Kiswandhi11} to
$2.4 - 2.9\%$ \cite{Chang}.

\begin{figure*}[htbp]

\includegraphics[width=0.65\linewidth,angle=0]{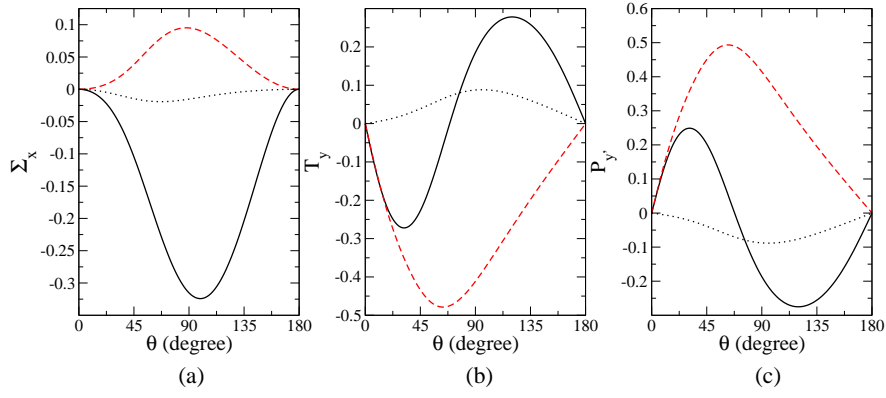}

\caption{(Color online) Single polarization observables for $\gamma p \to \phi p$ reaction: (a) polarized beam $\Sigma_x$, (b) polarized target $T_y$, and (c) recoil polarization $P_{y'}$ asymmetries, taken at photon laboratory energy $E_\gamma = 2.0$ GeV. The dotted lines denote the background contribution, while the solid black and red lines are contributions from resonances with $J^P = 3/2^-$ and $J^P = 3/2^+$, respectively.} \label{Polarization}
\end{figure*}

\begin{figure*}[htbp]

\includegraphics[width=0.65\linewidth,angle=0]{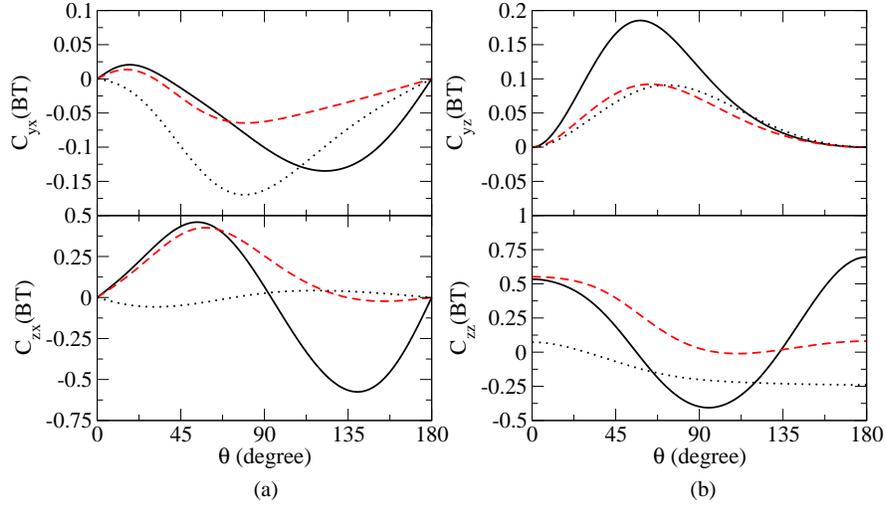}

\caption{(Color online) Double polarization observables for $\gamma p \to \phi p$ reaction: beam-target (BT) asymmetries (a) $C_{yx}^{BT}$ and $C_{zx}^{BT}$, (b) $C_{yz}^{BT}$ and $C_{zz}^{BT}$, with photon beam and nucleon target polarized, taken at photon laboratory energy $E_\gamma = 2.0$ GeV. Notation is as in Fig.\ref{Polarization}.} \label{BT}
\end{figure*}

\subsection{Polarization observables}

Since the fitted results with both assignments $3/2^\pm$ for the
resonance are rather similar, we need to find some observables that
can help us to distinguish them. Here, we show some predictions for
the polarization observables   at photon laboratory energy $E_\gamma
= 2.0$ GeV near the resonance position. Three single polarization
observables: asymmetries of the polarized beam $\Sigma_x$, polarized
target $T_y$, and recoil polarization $P_{y'}$ are given in Fig.
\ref{Polarization} while four double polarization observables:
beam-target (BT) asymmetries $C_{yx}^{BT}$, $C_{yz}^{BT}$,
$C_{zx}^{BT}$, and $C_{zz}^{BT}$, with the photon beam and the
nucleon target polarized, are given in Fig.~\ref{BT}, where the
dotted, solid, and dashed lines correspond to the contributions from
background only, and with the addition of the postulated resonance
of $J^P=3/2^-$ and  $J^P=3/2^+$, respectively. The notation of the
polarization observables follows Ref.~\cite{titov98}.

It can be concluded from Figs. \ref{Polarization} and \ref{BT} that while all the observables presented are reasonably distinct enough to distinguish the parities of the
$J = 3/2$ resonances, the single polarization observable $\Sigma_x$ is actually the most distinct based on the opposite sign of the curves produced by the two parities.

\section{summary and conclusions}
In summary, we present the details and  more extensive results of
the analysis of the near-threshold bump structure in the forward
differential cross section of the $\phi$-meson photoproduction to
determine whether it is a signature of a resonance.  The analysis is
carried out with an effective Lagrangian approach which includes
Pomeron and $(\pi, \eta)$ exchanges in the $t$ channel and
contributions from the $s$- and $u$-channel excitation of a
postulated resonance.

Besides the differential cross sections at forward angle as function
of photon energy and as function of $t$, the recent data on nine
spin-density matrix elements at three photon energies reported by
the LEPS collaboration are used, instead of the decay angular
distributions of the $\phi$ meson, which depend only on six
spin-density matrix elements, as was done before in
Ref. \cite{Kiswandhi10}, to constrain the model. Moreover, the new
spin-density matrix element data are given as a function of $t$,
while the previous decay angular distribution data are not.
Therefore, the new  set of data are expected to give more strict
constraints on the resulting resonance parameters.

We conclude that indeed the nonmonotonic behavior, along with the other experimental data,
as reported by LEPS, can only be explained with an assumption of the
excitation of a resonance of spin 3/2, as previously reported.
However, both parities of $(\pm)$ can account for the data equally
well with almost identical mass of $2.08\pm 0.04$ GeV and width of $
0.501\pm 0.117$ and  $0.570\pm 0.159$ for $3/2^-$ and $3/2^+$,
respectively. Spin-1/2 resonances can still explain the nonmonotonic behavior, but however would
lead to large resonance contributions, which would cause differential cross sections as functions of $t$, as well as
the spin-density matrix elements to disagree with experimental data.

The  helicity amplitudes of the $J^P = 3/2^-$ resonance calculated
from the obtained coupling constants gives a ratio of
$A_{1/2}/A_{3/2} = 1.05$ which  differs in sign from the value of
$-1.18$ of $D_{13}(2080)$ given by the PDG. Therefore, we conclude
that the $J^P = 3/2^-$ resonance cannot be identified as
$D_{13}(2080)$. The ratio of helicity amplitudes of the $J^P =
3/2^+$ resonance is obtained to be of $A_{1/2}/A_{3/2}=0.89$.

Some of the single and double polarization observables which are
sensitive to the parity of the resonance, including beam asymmetry
$\Sigma_x$, target asymmetry $T_x$, recoil asymmetry $P_x$, and
beam-target  asymmetry $C_{ij}^{BT}$, near the resonance peak are
also given.   Measurement of these quantities would be most helpful
in further substantiating whether the nonmonotonic behavior is
indeed a signature of resonance as well as resolving   its parity.
We find that the single polarization observable of beam asymmetry
$\Sigma_x$ provides an excellent way to resolve the parity of the
resonance since they are of opposite signs with different parity.

We have also investigated the effects of the postulated resonances
to the differential cross section of $\omega$ photoproduction as a
function of $t$ within the model of Ref. \cite{oh02}. We find that the
proposed resonance improves the agreement with the data, especially
around the photon LAB energy of 2.1 GeV, if large values of
OZI-evading parameter $x_{\textrm{OZI}} = 12$ and $x_{\textrm{OZI}}
= 9$ for $J^P = 3/2^-$ and $J^P = 3/2^+$ resonances, respectively,
are assumed. Here, again, both resonances are equally capable of
improving the discrepancy between the data and the predictions of
Ref. \cite{oh02}. It adds support for the  resonance we postulate. We
argue that the large values of OZI-evading parameter
$x_{\textrm{OZI}}$ found imply that the postulated resonance might
contain a strangeness content of $P_s = 0.1 \sim 15\%$. If the
postulated resonance contains considerable amount of strangeness,
then it could couple  strongly to, say, $K\Lambda$ channel. Question
would then arise on how the coupled-channel effects would modify the
low-energy behavior of the nonresonant amplitude employed in this
investigation. This can be answered only with a full coupled-channel
calculations.

\begin{acknowledgments}
We would like to thank Drs. W.C. Chang,   A.I. Titov, T.-S.H. Lee,
and Yongseok Oh for useful discussions and/or correspondences. This
work was supported in parts by National Science Council of the
Republic of China (Taiwan) under grant NSC100-2112-M002-012. We
would also like to acknowledge the help from National Taiwan
University High-Performance Computing Center in providing us with a
fast and dependable computation environment which is essential in
carrying out this work.
\end{acknowledgments}

\end{document}